\documentclass{article}

\usepackage{xcolor}
\usepackage{arxiv}
\usepackage{placeins}
\usepackage[utf8]{inputenc} 
\usepackage[T1]{fontenc}    
\usepackage{hyperref}       
\usepackage{url}            
\usepackage{booktabs}       
\usepackage{amsfonts}       
\usepackage{nicefrac}       
\usepackage{microtype}      
\usepackage{lipsum}		
\usepackage{graphicx}
\usepackage{natbib}
\usepackage{doi}
\usepackage{amsmath} 

\title{Unveiling Political Influence Through Social Media: Network and Causal Dynamics in the 2022 French Presidential Election}

\author{ \href{https://orcid.org/0000-0001-9731-9381}{\includegraphics[scale=0.06]{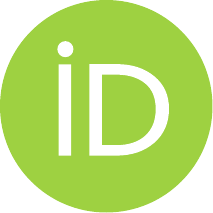}\hspace{1mm}Ixandra Achitouv}
        \\
	ISC-PIF - Institut des Systèmes Complexes - Paris Ile-de-France, CNRS\\
	\texttt{ixandra.achitouv@cnrs.fr} \\
	\And
	\href{https://orcid.org/0000-0001-9485-1399}{\includegraphics[scale=0.06]{orcid.pdf}\hspace{1mm}David Chavalarias} \\
	ISC-PIF - Institut des Systèmes Complexes - Paris Ile-de-France, CNRS\\
    CAMS - Centre d'Analyse et de Mathématique sociales 
}

\renewcommand{\shorttitle}{\textit{arXiv} Template}

\begin{document}
\maketitle

\begin{abstract}

    During the 2022 French presidential election, we collected daily Twitter messages on key topics posted by political candidates and their close networks. Using a data-driven approach, we analyze interactions among political parties, identifying central topics that shape the landscape of political debate. Moving beyond traditional correlation analyses, we apply a causal inference technique: Convergent Cross Mapping, to uncover directional influences among political communities, revealing how some parties are more likely to initiate changes in activity while others tend to respond. This approach allows us to distinguish true influence from mere correlation, highlighting asymmetric relationships and hidden  dynamics within the social media political network. Our findings demonstrate how specific issues, such as health and foreign policy, act as catalysts for cross-party influence, particularly during critical election phases. These insights provide a novel framework for understanding political discourse dynamics and have practical implications for campaign strategists and media analysts seeking to monitor and respond to shifts in political influence in real time.

\end{abstract}

\keywords{complex networks \and causal analysis \and computational social sciences}

\section{Introduction}

The rapid growth of social media platforms has transformed political campaigns, offering candidates tools to communicate, mobilize, and influence voters in real time. Consequently, understanding the dynamics of political interactions on platforms like Twitter-X has become a critical area of study. Prior research has explored social media networks in political contexts, with studies such as \cite{conover2011political,barbera2015birds} analyzing Twitter networks during elections to reveal patterns of polarization, ideological segregation, and influential actors. In \cite{larsson2017}, they analyzed the social media networks of political candidates during the Norwegian elections, revealing complex patterns of homophily and interaction among candidates on Twitter. Similarly, \cite{jones2019} investigated how digital architectures across platforms like Facebook and Twitter influence political campaigning strategies, underscoring the intricate relationship between platform affordances and political communication. Furthermore, studies such as those by \cite{castellano2020} have combined retweet networks with agent-based modeling to explore public discourse and polarization in political contexts, demonstrating how emergent phenomena arise from interactions within social networks.

In this study, we analyze the social media activity of political candidates and their networks during the French 2022 presidential election, focusing on 11 selected topics. Activity is defined as the daily number of tweets each political community has posted. As a result, we have 11 topics times 7 political communities, resulting in 88 time series over a period of 419 days.
The interplay between these political communities and topics exhibits properties of a complex system, characterized by interdependent dynamics, emergent patterns of activity, and non-linear temporal evolution. In this article, we characterize the dynamics of the political interactions, using network based approaches, spectral analysis and causal relationships test, offering new insights into the dynamics of political discourse and interaction.

Network-based approaches have been used to uncover hidden relationships, as seen in \cite{bialek2001complexity,himelboim2013social}, who used centrality metrics to identify key influencers. Furthermore, collective behaviors in time-varying correlation matrix have been studied through spectral analysis, where the largest eigenvalue of correlation matrices signals synchronization or coordination \cite{mcgraw2005eigenvalue,weng2013virality}. Causal inference techniques, have further been employed to explore directional influences between entities, as demonstrated by \cite{klinger2015social}.

This paper extends these work investigating the interactions between political communities during the 2022 French presidential election using Twitter data. The article is structured as follows: Section~\ref{sec:data} presents the dataset; Section~\ref{sec:collective} analyzes collective behavior through variations in party activity; Section~\ref{sec:corr} explores how network-based centrality measures reveal key community x topic nodes within the political landscape; Section~\ref{sec:causality} examines inter-party influence, focusing on how one community’s activity may affect another’s tweeting patterns. Finally, Section~\ref{sec:conclu} concludes the study. The data and codes used for this analysis will be available on \url{https://github.com/IxandraAchitouv/Dynamical_influence.git}

\section{Data collection}\label{sec:data}

The Politoscope is an observatory of the French political twittersphere that has been active between 2016 and 2023 recording political exchanges at the scale of the country \citep{gaumont_reconstruction_2018}. 
\begin{figure}[!ht]
    \centering
    \includegraphics[width=0.5\linewidth]{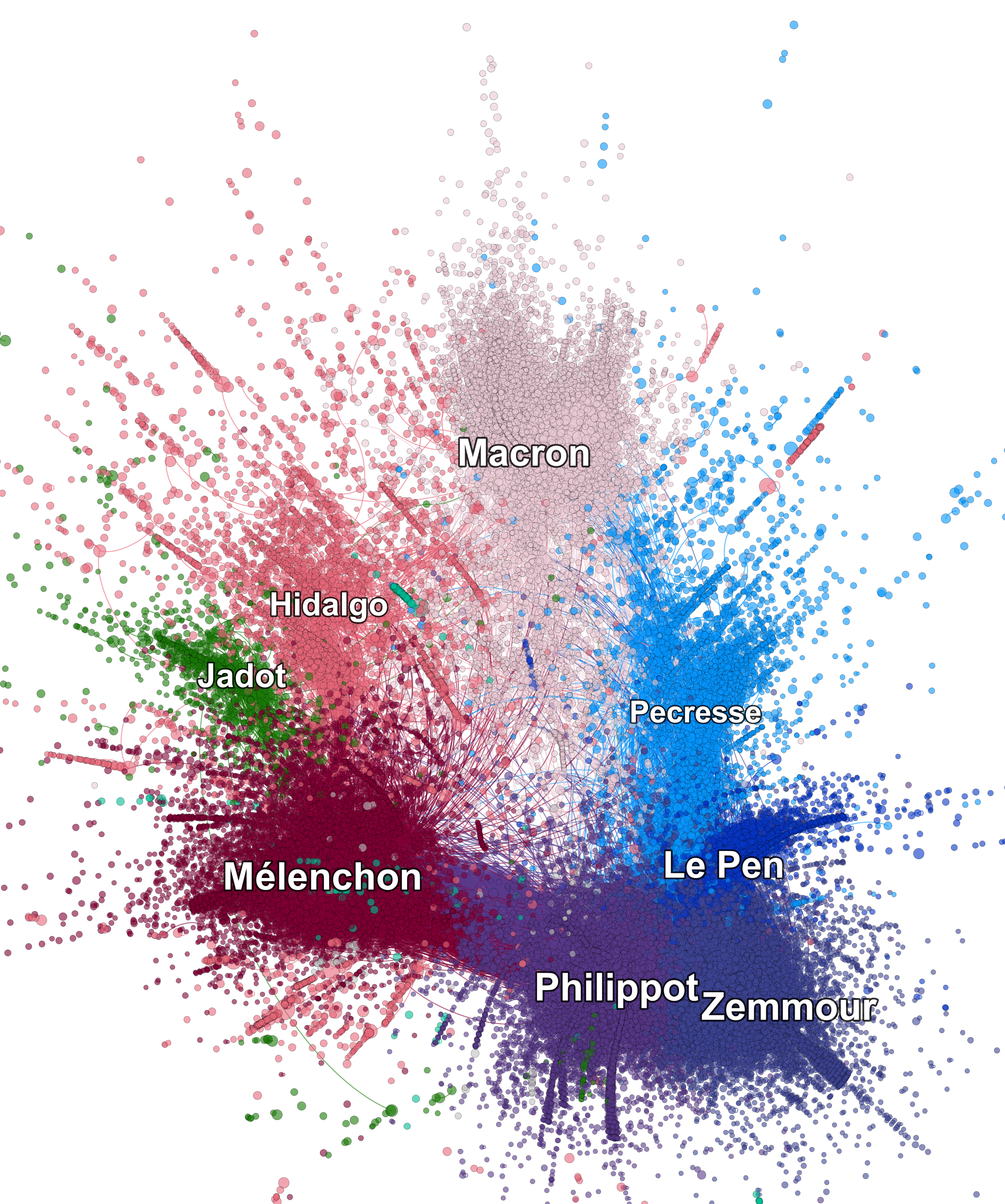}
    \caption{Political twittersphere over the period from 2021-09-01 to 2022-12-31. Following \cite{gaumont_reconstruction_2018}. Each node is a Twitter account, each color correspond to a political digital community. Political communities have been reconstructed from the retweets graph were links represent the retweet activity over the full period. Louvain clustering \cite{blondel_fast_2008} has been used to detect the communities and the graph has been spatialized with the Force Atlas algorithm \cite{jacomy2014forceatlas2}. Only links with weight highers than 20 retweets have been kept for the community detection and only links corresponding to 100 retweets or more are shown. Moreover, only the communities associated to political leaders studied in this paper are displayed on this image, which represent 39.4k accounts (80 \% of the full map). }
    \label{fig:Xnetwork}
\end{figure}

For doing so, it used the Twitte follow API to get the Twitter activity of about 3700 key political figures and used the Twitter track API to get the tweets mentioning some specific keywords. These keywords have been determined by a text-mining analysis using two types of sources: policy measures reported in campaign programs, written in developed language, and candidate tweets, written in frequently spoken language. \cite{gaumont_reconstruction_2018} have demonstrated that this approach collects between 30 \% and 80 \% of the tweets related to candidates or selected terms.

The Politoscope computes the core political communities on a daily basis based on the retweet graph of the 15 previous days. A core community captures not only institutional actors (e.g., politicians, party accounts), but also highly active users embedded within political discussions. The method is based on filtering the retweet graph to retain users who have mutually interacted through retweets at least w times over a 14-day window. Specifically, an edge is formed between two users if there are at least three retweets in either direction between them during that period. Moreover, to be included in the core, a user must be connected to at least three other users via such links—i.e., with at least three distinct neighbors for whom the mutual retweet count exceeds the threshold. This dual constraint ensures that the core consists of users who are both active and well embedded within the political conversation, excluding peripheral accounts or one-off interactions. The resulting subgraph represents a densely connected and politically engaged subset of the broader network, suitable for analyzing community structure, discourse dynamics, and influence. Then, communities are then labeled by the political leaders they include and this label is propagated to all the tweets of the accounts of the leaders's community. The precision in attribution of ideological alignment with this method has been shown to be greater than 90 \% \citep{gaumont_reconstruction_2018}.

This paper leverages the data collected by the Politoscope during the period 2021-09-01 to 2022-12-31 and the associated community attributions: 288M tweets involving 8.4M retweeting accounts and 11M accounts producing original content. Among these 288M tweets, 24.89\% are original tweets, 57.63\% are retweets, the rest being comments on tweets or retweets with comments. We also relied on the analysis of \citep{gaumont_reconstruction_2018} to define the main broad topics of the political debate during the covered period. With the help of a semantic map generated from the 2017 presidential programs and the Twitter corpora, eleven topics have been identified as particularly important and representative of the French political debate. These topics are described groups of terms obtained with text-mining methods and qualitative analysis of the political tweets. For this paper, we have updated the original topics of \citep{gaumont_reconstruction_2018} through marginal additions of terms that have been central in the political debate over the targeted period. The list of terms associated to each query can be found in the Appendix.  The objective was to obtain a query that makes it possible to extract from the \textit{Politoscope} database a set of coherent tweets related to each topics. Topics have been labeled by one of their most generic term for reference in the visualisation: agriculture, democracy, economics, ecology, foreign politics, immigration, employment, homeland security, research and education, health, energy policy. In Fig.\ref{fig:Xnetwork} we display a sample of the core political communities who produced the tweets related to our analysis.

\section{Preliminary analysis}

\begin{figure}[!h]
    \centering
    \includegraphics[width=1\linewidth]{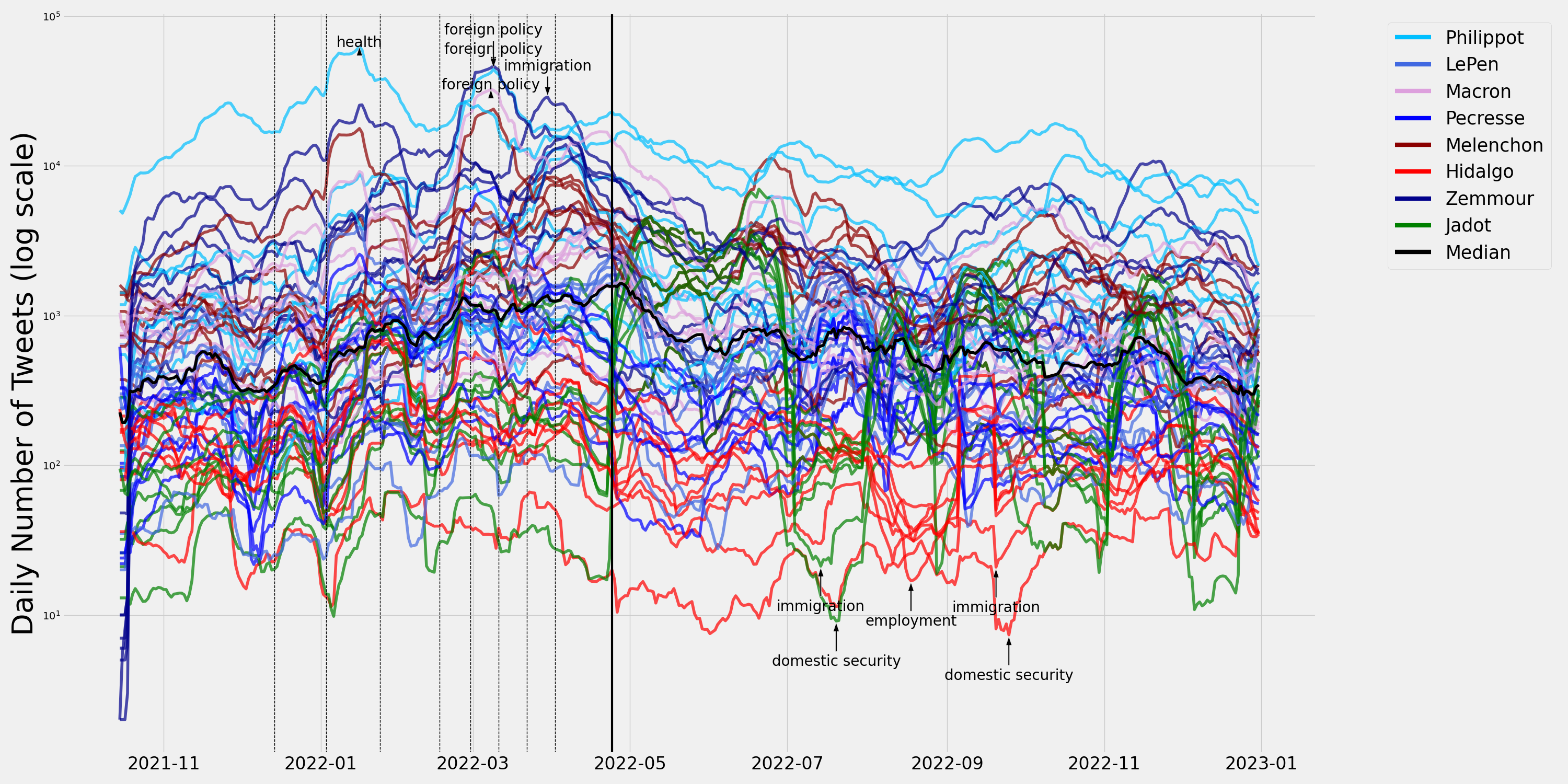}
    \caption{Time series of the daily number of tweets } 
    \label{fig:fig0}
\end{figure}

Using the key topics and the selected height key political figures with their core communities, we end up with 88 times series of 419 days long. Fig.\ref{fig:fig0} displays these time series (number of daily tweets per community per topic). The y-axis is in log-scale, revealing a large disparity among the activity between the different political parties. The vertical solid black line corresponds to the presidential election. Annotations show the maximum/lowest top 5 of these activities. Among the most active parties on Twitter we find Philippot and Zemmour on the topics health, foreign policy and immigration
while the less active communities are Hidalgo and Jadot on topics domestic security, employment, and immigration. The black curve corresponds to the median of all time series. As expected, the median number of tweets increases until the election and then decreases.   

\begin{figure}[!h]
    \centering
    \includegraphics[width=1\linewidth]{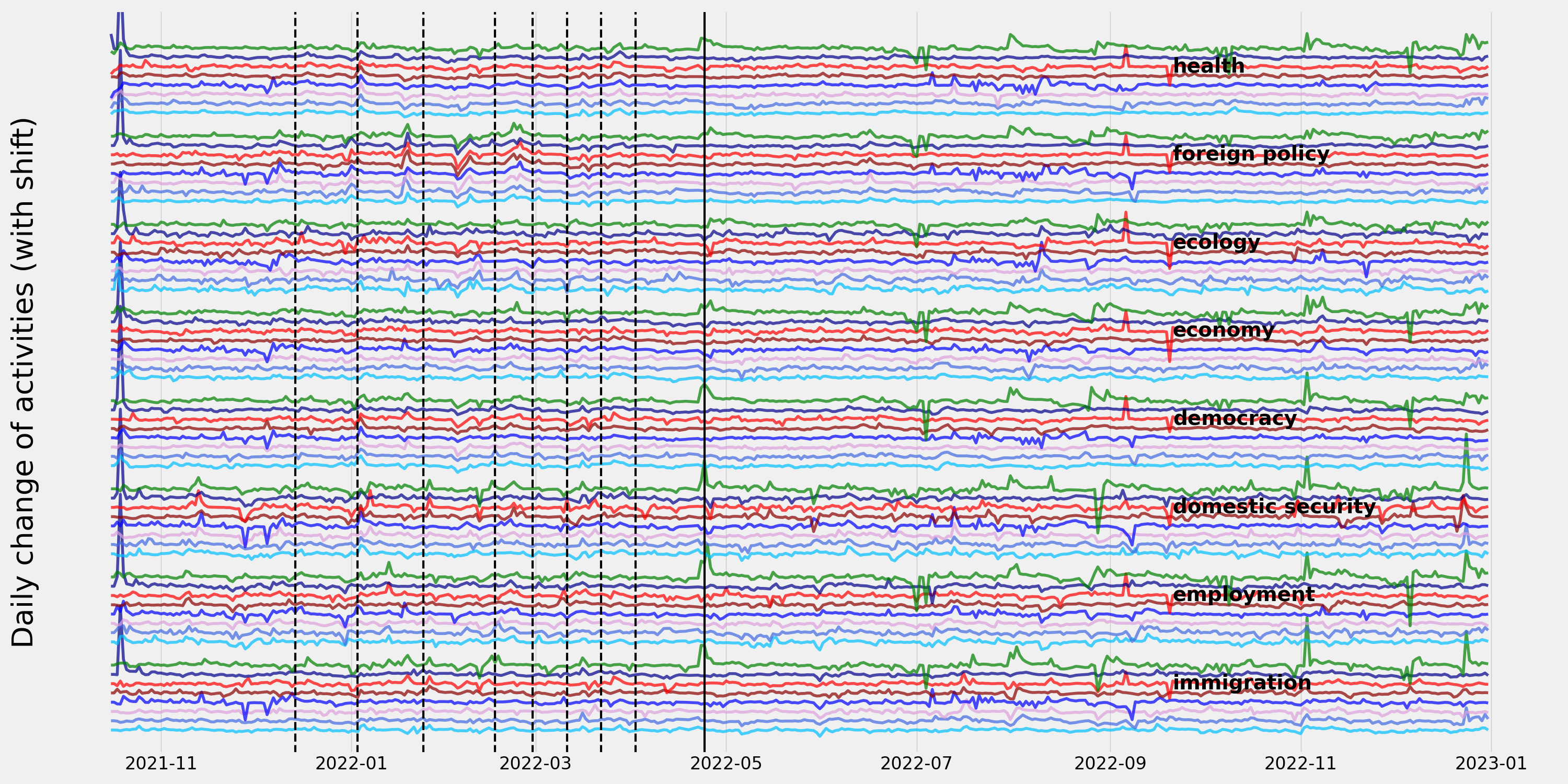}
    \caption{Time series of log-return of number of tweets (change of activities). Each time series is shifted to avoid overlapping and regrouped by topics for clarity (they all have a mean of zero).  } 
    \label{fig:fig0b}
\end{figure}
Fig.\ref{fig:fig0b} corresponds to the logarithm difference (log-returns) of the number of tweets on two consecutive days which is referred as the change of activities. Each of these times series are stochastic with a mean of zero but we shift these time series along the y-axis for better visibility. Qualitatively, we see that the change in activity per party seems to be more correlated than between parties. 

The vertical dashed lines in these two panels correspond to changes in the collective dynamics as defined in sec.\ref{sec:collective}. Is it noteworthy that most changes in the collective dynamics occur before the election.

\section{Detecting collective behavior and central nodes}\label{sec:collective}

In many complex systems, such as financial markets, where log-returns of stock prices are analyzed, or neuronal networks, where neural firing rates or membrane potentials are measured, the activity of individual entities exhibits time-dependent fluctuations that can be modeled as stochastic time series. These typically fluctuate around a zero mean, with no obvious deterministic trend. Despite this apparent randomness, correlated patterns and collective behavior can emerge, revealing underlying structure and coordination. 

Such collective behavior can be identified by analyzing the eigenvalue spectrum of the correlation matrix derived from the system’s time series. Collective dynamics often manifest itself as large eigenvalues, corresponding to strongly correlated modes that reflect coordinated behavior, synchronization, or critical transitions. These modes have been observed across domains, including systemic market movements in finance \cite{Plerou2002} and synchronized neural activity in brain networks \cite{Singh2020}. Monitoring temporal changes in the largest eigenvalues provides a quantitative approach for detecting shifts in system behavior, such as the onset of emergent phenomena or global coordination.

The emergence of large eigenvalues still holds as an indicator of collective dynamics, even if the entire eigenvalue distribution does not follow the Marchenko-Pastur law, as shown in studies that explore deviations from random matrix theory in systems with long-range correlations or complex internal structures \cite{Sornette2009}.

To detect global coordination in the change of activities of our political communities, we compute the correlation matrix over a sliding time window of \( \Delta t = 14 \) days across the entire time period \( T = 419 \). The correlation matrix for each pair of time series \( \xi_i(\Delta t) \) and \( \xi_j(\Delta t) \) is calculated as:

\begin{equation}
\text{Corr}_{ij}(\Delta t) = \frac{\langle \xi_i(\Delta t) \xi_j(\Delta t) \rangle - \langle \xi_i(\Delta t) \rangle \langle \xi_j(\Delta t) \rangle}{\sqrt{\langle \xi_i^2(\Delta t) \rangle - \langle \xi_i(\Delta t) \rangle^2} \cdot \sqrt{\langle \xi_j^2(\Delta t) \rangle - \langle \xi_j(\Delta t) \rangle^2}}\label{eq:corr}
\end{equation}

where $ \langle \xi_i(\Delta t) \rangle$  represents the mean of $ \xi_i(\Delta t)$  over the 14-day window, and $\langle \xi_i^2(\Delta t) \rangle$  is the mean of the squared values of $ \xi_i(\Delta t)$. The correlation matrix computed for each $ \Delta t $ is then analyzed by performing an eigenvalue decomposition, where the largest eigenvalue $ \lambda^{max}_{\Delta t} $ is computed. To detect significant collective dynamics, we require that the largest eigenvalue computed on the sliding window $ \lambda^{max}_{\Delta t} $ increases by an order of magnitude compared to the largest eigenvalue $ \lambda^{max}_{T }$ computed over the entire time period $ T $:

\begin{equation}
\lambda^{max}_{\Delta t}/\lambda^{max}_{T}-1 \geq 1 \label{eqcondi}
\end{equation}

In fig.\ref{fig:fig0}, we saw that the logarithmic return of the daily number of tweets were stationary time series with spikes,  similar to behaviors observed in complex systems where abrupt events disrupt otherwise stable patterns \cite{mandelbrot2004}. Similar properties have been observed in political social media activity, where sudden bursts of communication often correspond to significant political events or controversies \cite{castellano2020}. The vertical dashed lines correspond to Eq.\ref{eqcondi} being satisfied. Interestingly, we see that large increases of collective behavior happen during the presidential campaign. Spectral analysis therefore enables us to quantitatively characterize the period during which political communities are in competition, which a simple qualitative analysis of the logarithmic return of the daily number of tweets could not reveal.

To study the origins of these synchronized changes within political parties, we used network analysis. We start by averaging our time series $\xi$(political community x topic) over the political parties and then we compute the correlation matrix among these 11 (topics) time series. In order to map the correlation matrix computed at each time step to an adjacency matrix we use the threshold method (e.g. \cite{Newman2010, Watts1998,achitouv2024dynamicalanalysisfinancialstocks}). The threshold method involves setting a threshold $\rho_c$ such that only the pairs of nodes with correlation coefficients greater than $\rho_c$ are considered connected. Mathematically, we define the adjacency matrix $A$ as:

\begin{equation}
A_{ij} = 
\begin{cases} 
|\text{Corr}_{ij}| & \text{if } |\text{Corr}_{ij}| \geq \rho_c \\
0 & \text{otherwise}\label{eq:filtering}
\end{cases}
\end{equation}

where $ \text{Corr}_{ij} $ is the correlation between time series \( \xi_i \) and \( \xi_j \), and $\rho_c$ is the threshold value. This adjacency matrix can then be used to construct a graph where nodes represent political parties associated with topics (or just topics), and edges represent the strength of the correlations. In what follows, we use as a threshold the 90th quantile of the distribution of absolute values in the correlation matrix\footnote{As it is discussed in \cite{achitouv2024dynamicalanalysisfinancialstocks}, there is no obvious criteria for the choice of the threshold, however one could take the minimum threshold that produces a degree distribution of the nodes similar to a power law for which the centrality measure tend to converge as noises are filtered out.}.

We then compute centrality measures on each graph. The weighted degree of a node is the sum of the weights of its connections to other nodes:

\begin{equation}
    C_{\text{degree}}(i) = \sum_{j} A_{ij}\label{eq:Cdeg}
\end{equation}

where $A_{ij}$ is the adjacency matrix of the network, and $C_{\text{degree}}(i)$ is the degree centrality of node $i$. This measure gives an indication of how strongly the dynamics of the node is correlated with the global dynamics.

We also measure the betweenness centrality in what follows. This quantity quantifies the extent to which a node is bridging different communities in the network. It is defined as:

\begin{equation}
    C_{\text{betweenness}}(i) = \sum_{s \neq i \neq t} \frac{\sigma_{st}(i)}{\sigma_{st}}
\end{equation}

where $\sigma_{st}$ is the number of shortest paths between nodes $s$ and $t$, and $\sigma_{st}(i)$ is the number of those paths that pass through node $i$. Nodes with high betweenness centrality play a crucial role in connecting different parts of the network,
acting as brokers or gatekeepers of information. In our case, those nodes can be interpreted as the speeches of certain candidates that cross various aspects of French political life.

These two metrics help us understand how political parties and topics are interrelated and make it possible to highlight key players and topics in the discourse during different phases of the election.

\begin{figure}[!h]
    \centering
\includegraphics[width=1\linewidth]{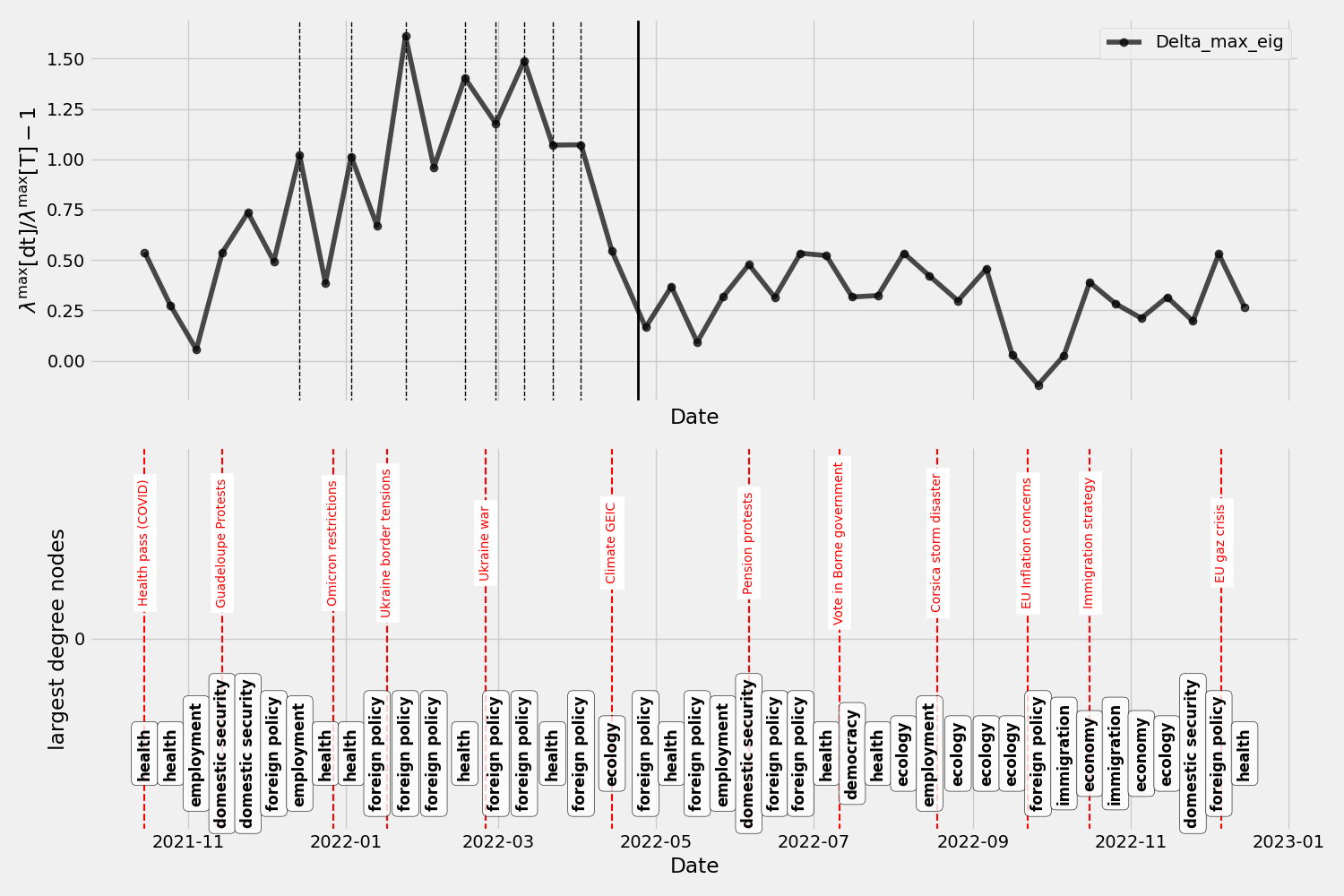}
    \caption{Top panel: Evolution of the largest eigenvalues relative to those computed over the entire period.
Bottom panel: Topics corresponding to the highest degree, annotated with key political events.}
    \label{Fig:collective}
\end{figure}

\begin{table}[!h]
\centering
\begin{tabular}{|p{3.5cm}|p{6cm}|p{4.5cm}|p{2.8cm}|}
\hline
\textbf{Event} & \textbf{Description} & \textbf{Source URL} & \textbf{Date} \\
\hline
COVID-19 & Covid health pass & \href{https://www.france24.com/en/europe/20211014-macron-s-covid-health-pass-a-success-in-overcoming-france-s-vaccine-scepticism}{France24} & 2021-10-14 \\
\hline
Guadeloupe Protests & Protests erupt over health pass and mandates & \href{https://www.france24.com/fr/france/20211121-nouvelle-nuit-de-violence-en-guadeloupe-en-proie-%C3%A0-un-mouvement-social-d-ampleur}{France24} & 2021-11-14 \\
\hline
Omicron restrictions & France restricts travel, updates health measures & \href{https://www.france24.com/en/europe/20211227-france-considers-new-covid-restrictions-ahead-of-expected-omicron-surge}{France24} & 2021-12-27 \\
\hline
Ukraine border & Diplomatic talks over Ukraine tension & \href{https://edition.cnn.com/2022/01/20/europe/ukraine-russia-tensions-explainer-cmd-intl}{CNN} & 2022-01-17 \\
\hline
Ukraine war & Russia invades Ukraine, France reacts diplomatically & \href{https://edition.cnn.com/europe/live-news/ukraine-russia-news-02-23-22}{CNN} & 2022-02-24 \\
\hline
Climate debate & GEIC repport and Presidential candidates debate ecology topics & \href{https://www.euronews.com/green/2022/04/21/how-france-s-environmental-policies-could-change-after-this-weekend-s-elections}{Euronews} & 2022-04-14 \\
\hline
Pension protests & Mass protests against proposed reforms & \href{https://www.france24.com/en/europe/20230606-%F0%9F%94%B4-live-france-faces-14th-day-of-nationwide-protests-against-pension-reform}{France24} & 2022-06-06 \\
\hline
Vote in Borne government & PM Elisabeth Borne wins confidence vote & \href{https://www.france24.com/en/france/20220711-macron-under-pressure-over-uber-leaks-french-pm-faces-no-confidence-vote}{France24} & 2022-07-11 \\
\hline
Corsica storm disaster & Violent storms cause deaths and damage in Corsica & \href{https://www.lemonde.fr/en/environment/article/2022/08/19/corsica-storm-people-were-screaming-everywhere-the-trees-fell-like-matchsticks_5994100_114.html}{Le Monde} & 2022-08-18 \\
\hline
Immigration strategy & Government proposes new immigration law & \href{https://www.france24.com/en/france/20241016-france-interior-minister-retailleau-charts-new-hardline-immigration-strategy}{France24} & 2022-10-06 \\
\hline
EU Inflation concerns & Fears of European companies relocating to the United States & \href{https://www.lemonde.fr/en/opinion/article/2023/09/22/the-inflation-reduction-act-does-not-present-major-economic-security-risks-for-the-eu_6138513_23.html}{Le Monde} & 2022-12-13 \\
\hline
EU gaz crisis & Reduction of Russian gaz in EU & \href{https://www.lemonde.fr/economie/article/2022/12/06/europe-la-crise-du-gaz-devrait-durer-au-moins-jusqu-en-2027_6153205_3234.html}{Le Monde} & 2022-12-06 \\
\hline
\end{tabular}
\caption{Events with media coverage related to our topics}
\label{tab:events}
\end{table}

In Fig.\ref{Fig:collective} top panel, we display the the ratio $\lambda^{max}_{\Delta t}/\lambda^{max}_{T}-1$ computed on our time period using the activities variation for $\xi_i (\Delta t)$ in Eq.\ref{eq:corr}. On the bottom panel we show the topic that had the highest centrality (Eq.\ref{eq:Cdeg}), along with events reported in media (see Tab.\ref{tab:events})

\begin{table}[]
\centering
\begin{tabular}{|l|l|l|l|}
\hline
\textbf{Topic} & \textbf{Electoral campaign (18w)} & \textbf{Post election (24w)} & \textbf{Total} \\ \hline
\textbf{Foreign policy}         & 7 & 6 & 13 \\ \hline
\textbf{Health}                 & 6 & 4 & 10 \\ \hline
\textbf{Domestic security}      & 2 & 2 & 4  \\ \hline
\textbf{Employment}             & 2 & 2 & 4  \\ \hline
\textbf{Ecology}                & 1 & 5 & 6  \\ \hline
\textbf{Immigration}            & 0 & 2 & 2  \\ \hline
\textbf{Democracy}              & 0 & 1 & 1  \\ \hline
\textbf{Economy}                & 0 & 2 & 2  \\ \hline
\textbf{Agriculture}            & 0 & 0 & 0  \\ \hline
\textbf{Research and Education} & 0 & 0 & 0  \\ \hline
\textbf{Energy policy}          & 0 & 0 & 0  \\ \hline
\end{tabular}\caption{Number of occurrences of topics as the highest degree topic in the correlation matrix before and after the election. \label{Htopics}}
\end{table}

Interestingly, the end of the election period not only corresponds to a global change in collective dynamics, but also to a change in the prevalence of the most central topics in the political landscape as measured by the correlation matrix through time (see table~\ref{Htopics}). In particular, ecology and environmental issues, which was the second major concern of the French (84\%) after purchasing power (91\%) is almost absent in the top pre-electoral topics while it was well represented after the election.

During the presidential campaign, some topics are over-represented compared to their post-election frequency:  health (related to health measures implemented following the arrival of the COVID-19 Omicron variant) and the rising tensions at the Ukraine border, followed by the outbreak of the Ukraine war (topics: health and foreign policy) while some others simply never appear: agriculture, research and education and energy policy.

This observation suggests that political communities put in place strong agenda-setting strategies, highlighting issues likely to give them an advantage in the vote. To further analyze the differences between the pre and post election activity of political communities, we analyze the graph constructed from Eq.~\ref{eq:filtering} before and after the election.

\section{Evolution of the Political Landscape}\label{sec:corr}

A comparison of the pre- and post-2022 French presidential election graphs (constructed from the correlations of daily variations) reveals significant shifts in the political landscape, as evidenced by changes in centrality measures.

\begin{figure}[h!]
    \centering
    \includegraphics[width=0.7\textwidth]{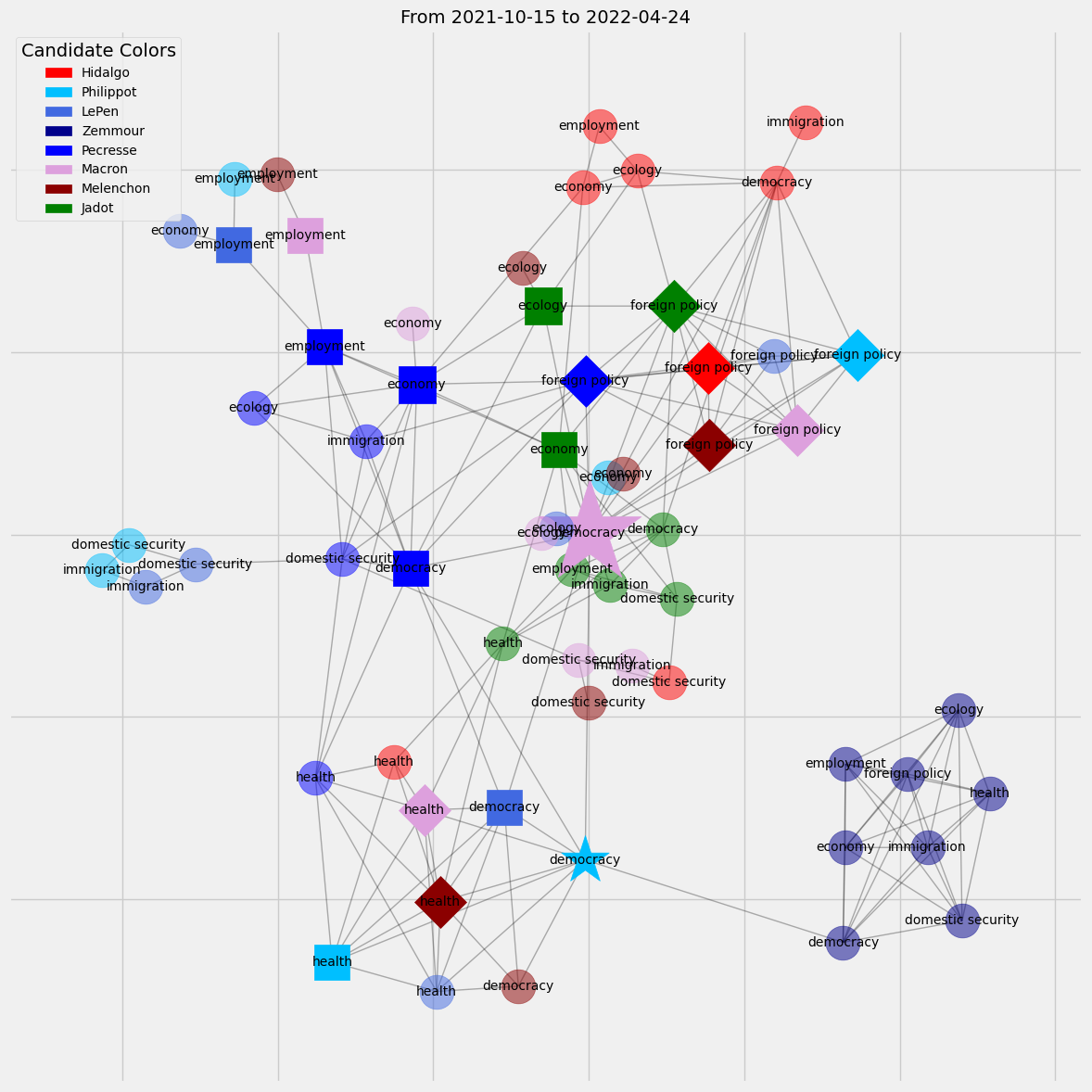}
    \caption{Correlation network structure pre-election}
    \label{fig:figure1}
\end{figure}
\begin{figure}[h!]
    \centering
    \includegraphics[width=0.7\textwidth]{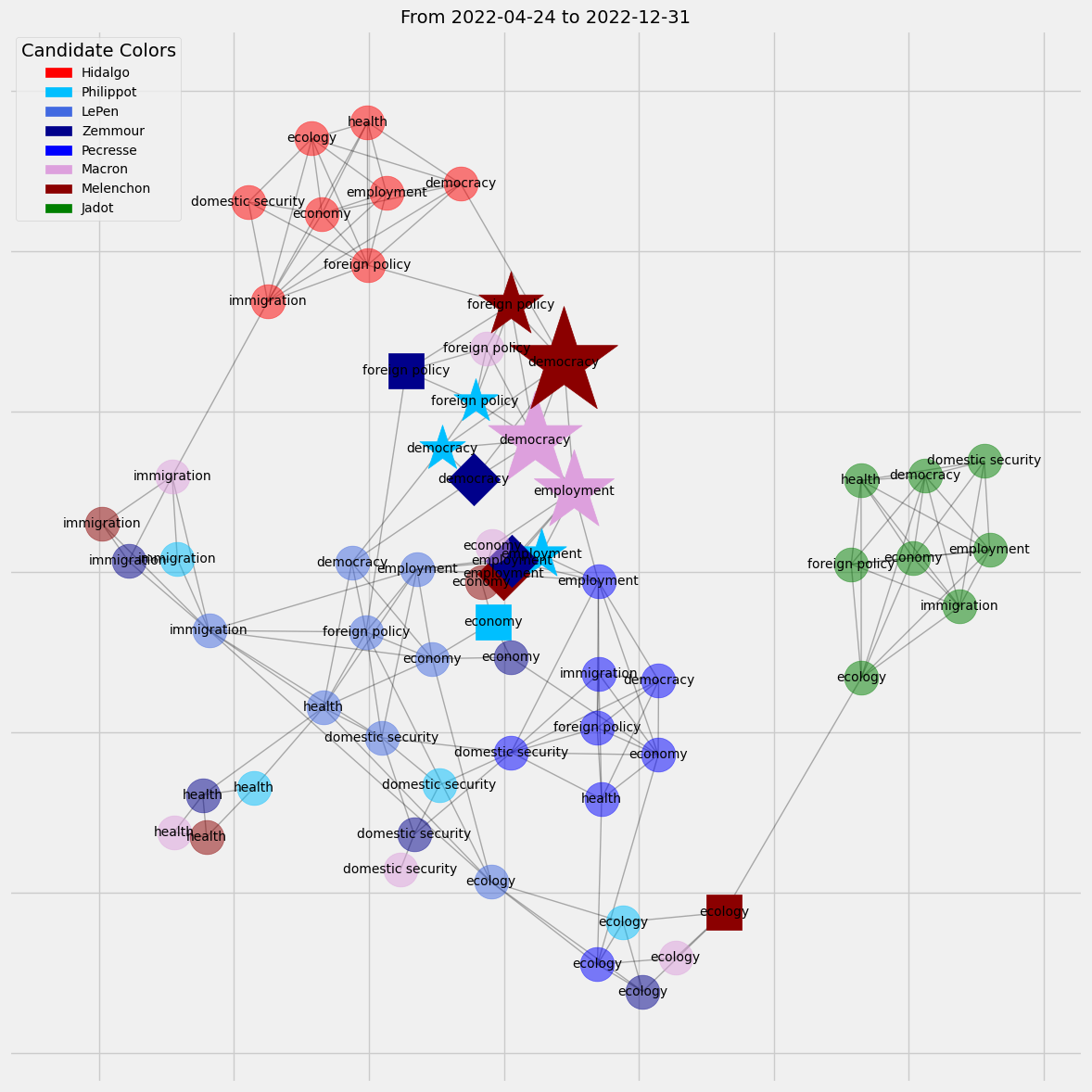}
    \caption{Correlation network structure post-election}
    \label{fig:figure2}
\end{figure}

In Fig.~\ref{fig:figure1} and \ref{fig:figure2}, the resulting graphs are shown, where colors represent political communities. Nodes depicted as squares have betweenness centrality in the top 25th percentile of the distribution, those shown as diamonds have degree centrality in the top 25th percentile, and nodes shown as stars are in the top 25th percentile for both betweenness and degree centrality distributions. The size of these nodes is proportional to their centrality measure. The spatial visualization corresponds to Force Atlas 2 \cite{jacomy2014forceatlas2}, which is a force-directed layout algorithm that simulates a physical system where nodes repel each other like charged particles and edges act as springs, attracting connected nodes together. This process results in a network layout where structural proximities are reflected as visual proximities. Interestingly, we observe some clustering by communities. In the pre-election period, we see that the political parties of Hidalgo and Jadot appear on the left, and Zemmour forms an almost distinct cluster on the right. Macron in the democracy topic is the central node both in terms of centrality and betweenness. We also see two clusters on foreign policy and on health where diverse parties (left and right) are connected.

After Macron's election, the political landscape changed. We see that now the Hidalgo community is a cluster connected to Mélenchon's node on democracy and foreign policy, while Jadot is also a separate cluster but connected to Mélenchon's community on the topic of ecology. In the post-election period, Mélenchon led efforts to unite the left for the legislative elections by forming the Nouvelle Union Populaire Écologique et Sociale (NUPES), an alliance of left-wing parties. This alliance aimed to make Mélenchon Prime Minister by winning a parliamentary majority. While it became the largest left-wing bloc in the National Assembly, it did not secure a majority. However, it explains how in this post-election period his communities became top nodes in both degree and betweenness centrality, alongside Macron, as he became the figurehead of the united left-wing political movement, making him a direct competitor to Macron with contrasting ideological views.

One should distinguish, however, between the interpretation of clustering or centrality of nodes among changes in activities and the influence of one community on another. In fact, these graphs are not directional, and one may wonder if a central node here is central due to its influence or because its political party is influenced by others. To test the influence of a node (political community × topic), we turn to causal inference.

\section{Causal inference as a measure of influence}\label{sec:causality}

Convergent Cross Mapping (CCM) is a powerful technique for detecting causality in complex, nonlinear systems. It was introduced by \cite{Sugihara2012} as a method for inferring causal relationships by examining how well the historical states of one time series can be used to predict the states of another. The CCM method goes beyond simple Granger causality inference \cite{granger1969investigating} as it is suited for nonlinear systems dominated by stochasticity, as is the case for the variation of activities.

The CCM method involves reconstructing the state space of each time series using delay embedding, as established by Takens’ Theorem \cite{takens1981detecting}. By constructing shadow manifolds for each series, CCM assesses whether information about one variable is recoverable from the reconstructed manifold of the other. If the cross-mapping skill (measured via prediction accuracy) increases as more data points are used, it suggests a causal relationship from one variable to the other.

Mathematically, the state-space reconstruction for a given time series \(X = \{x(t)\}\) is achieved through delay embedding, which creates a vector representation of the system's state at time \(t\), given by:

\[
M_{X}(t) = \big( x(t), x(t - \tau), x(t - 2\tau), \ldots, x(t - (E-1)\tau) \big)
\]

where \(\tau\) is the embedding delay and \(E\) is the embedding dimension. Similarly, for another time series \(Y = \{y(t)\}\), the reconstructed manifold is:

\[
M_{Y}(t) = \big( y(t), y(t - \tau), y(t - 2\tau), \ldots, y(t - (E-1)\tau) \big)
\]

To assess whether \(X\) causally influences \(Y\), Convergent Cross Mapping (CCM) evaluates whether the states of \(Y\) can be reconstructed using the manifold \(M_X\), which is built from the time-delay embedding of \(X\). The reconstructed estimate \(\hat{Y}(t)\) is obtained using the nearest neighbors of \(M_X(t)\) in the embedded space.

Let \(\{t_i\}_{i=1}^{E+1}\) be the time indices of the \(E+1\) nearest neighbors of the point \(M_X(t)\), found using Euclidean distances in the embedded space \(M_X\). The distances to these neighbors are denoted as 

\[
d(t, t_i) = \sqrt{ \sum_{k=0}^{E-1} \left[ x(t - k\tau) - x(t_i - k\tau) \right]^2 }
\]

and the weights are defined by:

\[
w_i = \exp\left(-\frac{d(t, t_i)}{\min_{1\le i\le E} d(t, t_i)}\right)
\]

\[
w_i = \exp\left( -\frac{d(t, t_i)}{ \displaystyle \min_{\substack{1 \le j \le E+1}} d(t, t_j) } \right)
\]

Using these weights, the cross-mapped estimate of \(Y(t)\), denoted \(\hat{Y}(t)\), is given by a weighted average of the corresponding values of \(Y\):

\[
\hat{Y}(t) = \sum_{i=1}^{E+1}\frac{ w_i}{\sum_{i=1}^{E+1} w_i}\, y(t_i)
\]

The accuracy of the reconstruction is evaluated by computing the Pearson correlation coefficient \(\rho\) between the actual values \(Y(t)\) and the estimates \(\hat{Y}(t)\):

\[
\rho = \text{cor}\big(Y(t), \hat{Y}(t)\big)
\]

An increasing \(\rho\) with more library points indicates that information about \(Y\) is encoded in \(M_X\), suggesting that \(X\) causally influences \(Y\).


The predictive skill is commonly measured by the Pearson correlation coefficient \(\rho\) between the observed values of \(X\) and their cross-mapped estimates from \(M_{Y}\). Formally, if \(\hat{X}(t)\) represents the estimates of \(X\) from \(M_{Y}\), then:

\begin{equation}
\rho = \text{cor} \big( X(t), \hat{X}(t) \big)    
\end{equation}\label{pearsoneq}

If \(\rho\) increases significantly as more data points are used in the reconstruction, then \(X\) is said to causally influence \(Y\). CCM measure has been applied to various fields, including to infer the non-linear influences among species in ecological systems \citep{Ye2016,Deyle2016,Mønster2017,Harford2017}.

In the context of analyzing political influence through Twitter data, CCM can be applied to our stationary time series (the variation in activities of the different political communities × topics).

Indeed, by determining whether the activity of one community causally influences another, we can move beyond correlation-based graphs to uncover directional relationships and assess how political influence propagates through the network. This approach can offer a clearer view of how various actors interact and influence each other’s discourse.

\subsection{Causal network}

\begin{figure}[!h]
    \centering
    \includegraphics[width=1\linewidth]{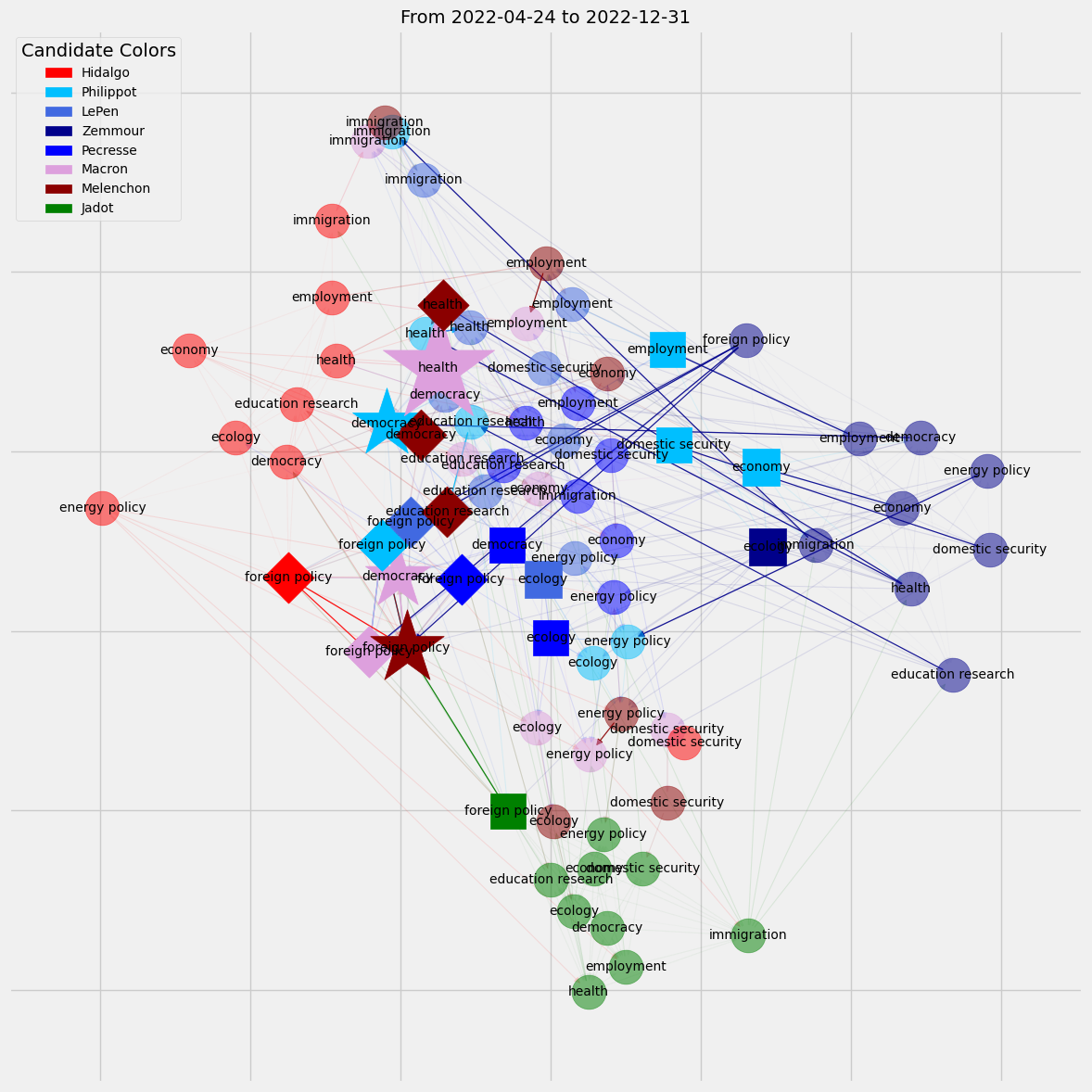}
    \caption{Causal network of the political communities x topics}
    \label{fig:causalnet}
\end{figure}

Figure~\ref{fig:causalnet} shows the network representation of causal influence. The edges indicate causal connections and their weights are proportional to the Pearson correlation coefficient from Eq.~\ref{pearsoneq}. We set the edge transparency based on the relationship: it is nearly invisible for influences within the same community, lightly transparent for cross-party influences with a Pearson coefficient below 0.5, and fully opaque when above that threshold. Node shapes and colors follow the same conventions as in the previous graphs.

Interestingly, we first observe clustering among the political communities of Hidalgo, Jadot, and Zemmour, while other communities appear more central. We also observe an interesting spatial distribution of the political communities according to their "left" or "right" political views: for instance, Philippot appears mostly between Zemmour and Le Pen/Pécresse. Beyond that, nodes representing Macron, Mélenchon, and finally Hidalgo. 

Among the strongest influences, we see that Jadot and Hidalgo influence Mélenchon on the topic of foreign policy. Hidalgo also influences Macron, but neither of them influences the right or extreme right political parties, nor are they influenced by them. While Zemmour’s cluster is not central in the graph, he has some strong influence on diverse parties, an effect that was not visible in the correlation-based graph.

With less significant influence (semi transparent edges), we note that foreign policy brought together political parties across the ideological spectrum, with candidates influencing one another beyond their left/right inclinations. From a sociological point of view, a robust democratic debate often thrives on the interaction of diverse viewpoints, as deliberative democracy theory suggests \citep{habermas1984theory,dryzek2000deliberative}. Consequently, it is pertinent to examine which topics acted as catalysts, prompting candidates to respond to one another and engage in a dynamic exchange. Investigating these dynamics may therefore shed light on the broader mechanisms that shape political dialogue and influence electoral outcomes.

\subsection{Political communities influence}
\begin{figure}[!h]
    \centering
    \includegraphics[width=0.8\linewidth]{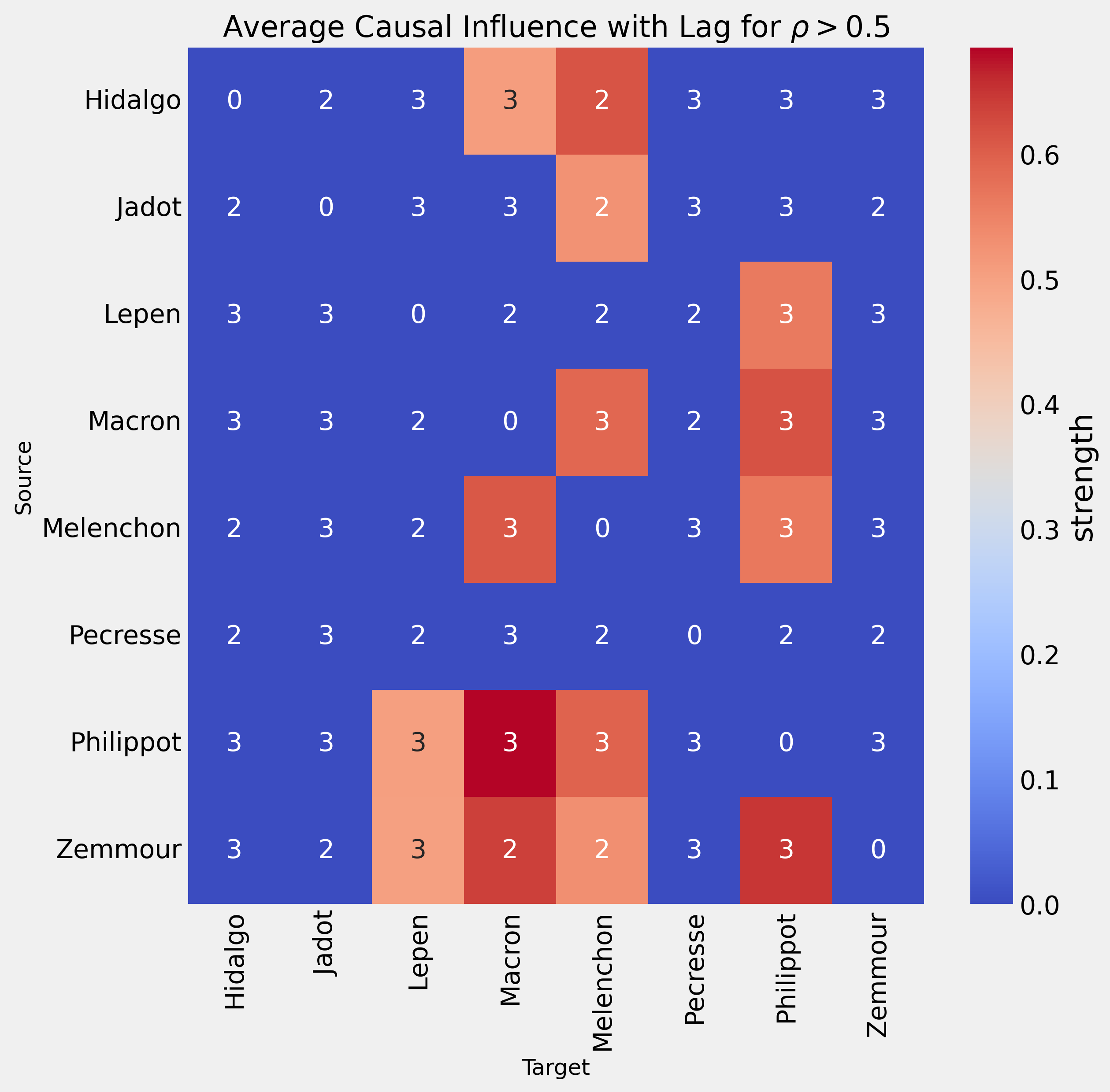}
    \caption{Matrix representation of the influence among political communities: rows correspond to sources, and columns correspond to targets. The threshold for influence is set to $\rho>0.5$}
    \label{fig:matrixinfluence}
\end{figure}

To quantify the overall influence of each political community, we aggregate the influence of each community averaged over all topics and display the results as a matrix, with the source of influence in the rows and the target in the columns, as shown in Fig.~\ref{fig:matrixinfluence}. Here, we only average meaningful causal inferences where the Pearson coefficient is $\rho>0.5$, masking other values to zero. Interestingly, we find again that the Hidalgo and Jadot communities neither influence right or extreme right parties nor are influenced by any community. Macron and Mélenchon parties influence one another, and Macron, Mélenchon, and Philippot are more influenced than other parties. Interestingly, the Zemmour party influences the most communities but is not influenced by anyone. The number associated with each entry of the matrix corresponds to the lag (t-lag) at which the strongest influence occurs, which here corresponds to t-3 or t-2 days while we test for lags up to t-7.

\begin{figure}[!h]
    \centering
    \includegraphics[width=1\linewidth]{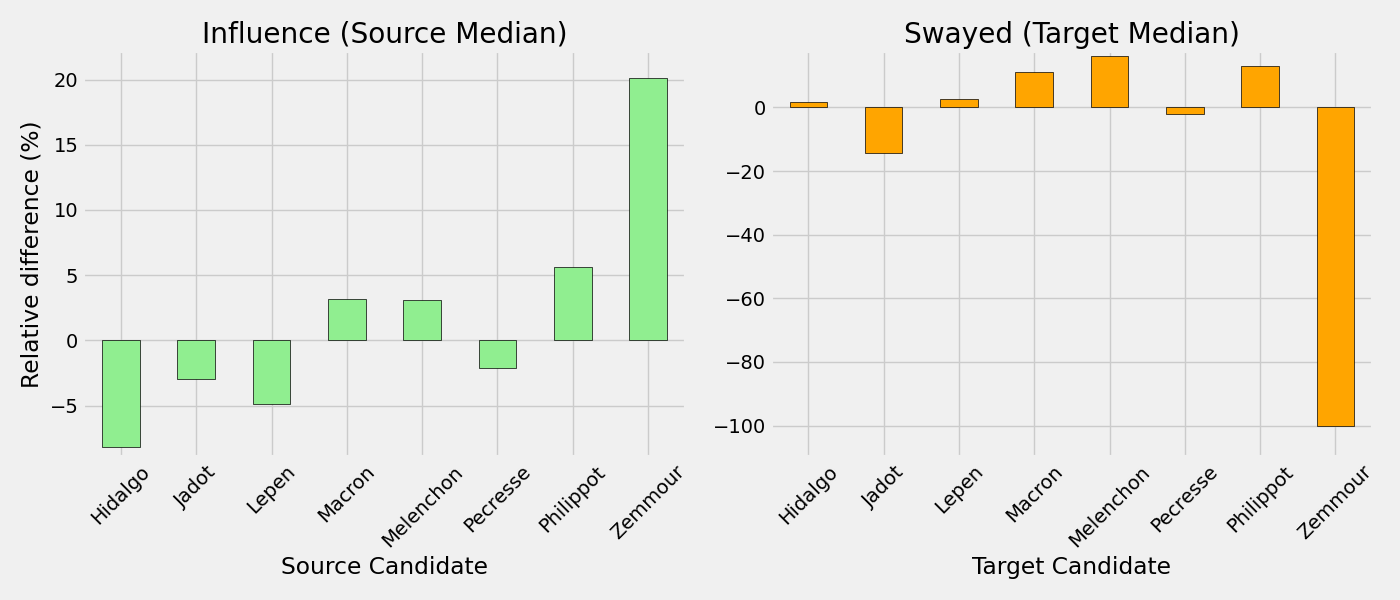}
    \caption{Relative influence of each party on all others is measured by the median (green bars). The orange bars represent how much each party is influenced by the others, relative to the median.}
    \label{fig:influence_allav}
\end{figure}

We can also average the influence (rows) over all political parties to determine which community influences the most or is most likely to be influenced (average per column). To test this, we consider a lower threshold for the Pearson coefficient, setting $\rho>0.25$, which corresponds to the 90th quantile of the overall influence matrix distribution (topics × political parties) and for which most entries are zero. We show this result in Fig.~\ref{fig:influence_allav} as relative influence compared to the median (left panel with green bars) and as the relative proportion of being influenced (swayed) compared to the median (orange bars) in the right panel. We see that Hidalgo, for instance, has the lowest influence, while Zemmour has the strongest. Macron, Mélenchon, and Philippot have positive influence compared to the median. On the other hand, the political communities that respond (are influenced) the most are Mélenchon’s, Philippot’s, and Macron’s, which is also reflected in the spatial distribution of the influence graph. Zemmour is a particular case: it is not influenced by other political communities but has a strong influence on them.

\subsection{Topics of influence}
\begin{figure}[!h]
    \centering
    \includegraphics[width=1\linewidth]{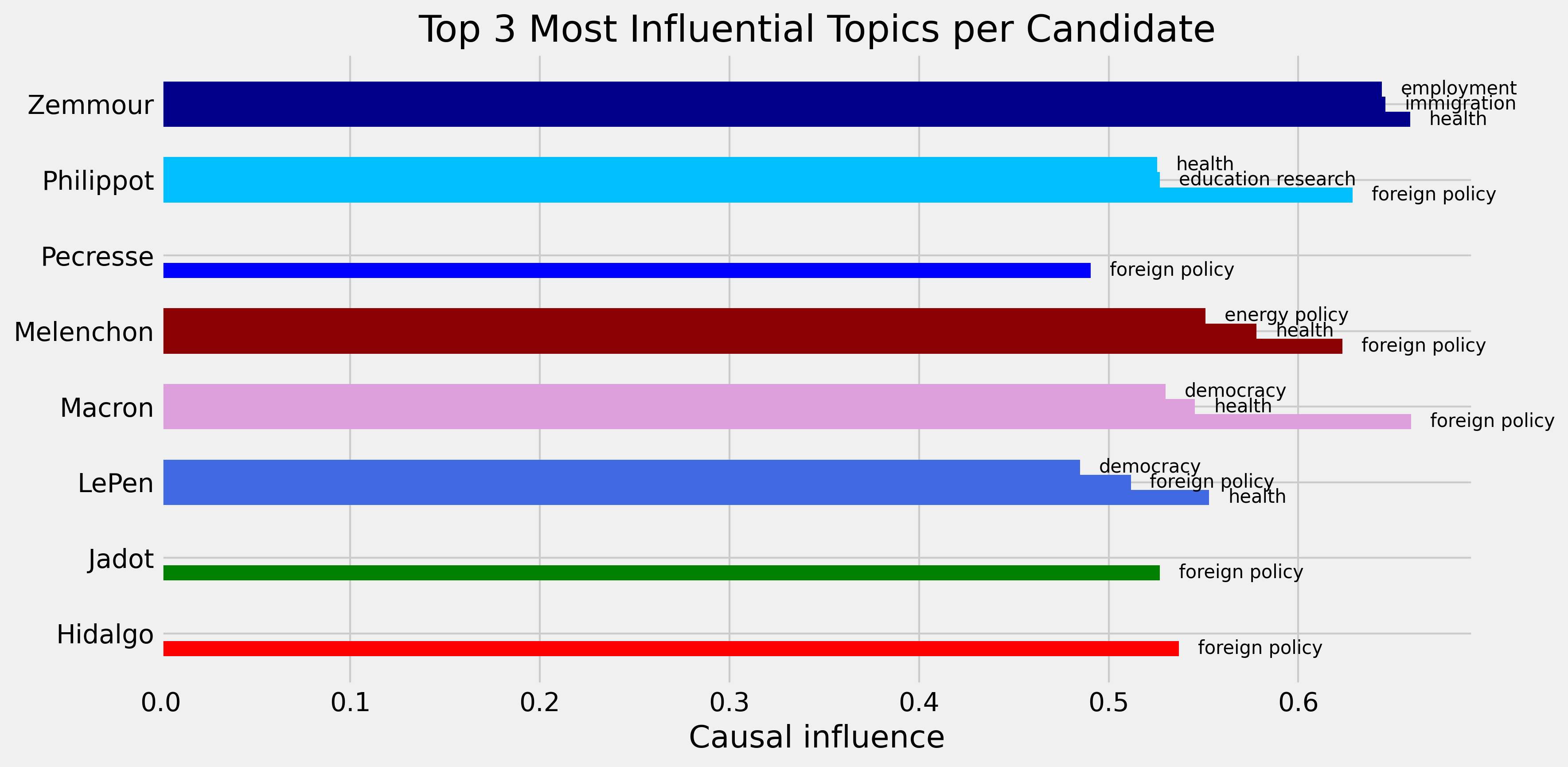}
    \caption{Top influential topics (max 3) per political community with threshold $\rho=0.4$.}
    \label{fig:top3}
\end{figure}

Finally, if we do not aggregate our measure of influence over all topics, we can also test which topic has the most influence per communities. For this we lower the threshold of influence $\rho>0.4$ in order to obtain a result for each political communities. As we can observe in Fig.\ref{fig:top3}, a threshold of $\rho>0.5$ would result in no influence for the community of Pecresse. In this figure we see the top 3 topics per political communities that influence. Unsurprisingly the topics that are the most influent are health (following the COVID pandemic and the preventive measures taken by the government), and foreign policy. The latter can be understood in light of the Russian invasion of Ukraine, which began on February 24, 2022. The war heightened concerns over Europe’s defense, security, and autonomy, prompting French political parties to shift their rhetoric on European sovereignty.

In addition, some political parties have no more than one topic influencing others with a threshold of $\rho>0.4$ (Pécresse, Jadot, and Hidalgo). Among Zemmour’s influential topics, immigration is part of the top three. During his campaign, Zemmour also advocated for stricter immigration policies, often emphasizing national identity and perceived threats of multiculturalism \citep{newsweek2022zemmour}. His rhetoric has positioned him as a key figure in the immigration debate, further amplifying his party’s influence. Interestingly, the most influential topics within the Le Pen and Macron communities are the same. To recall, Macron and Le Pen were competitors for the second round of the presidential election. However, we see that Macron’s influence is stronger than Le Pen’s, which is also supported by the correlation graph–based measures of centrality.

\section{Conclusion}\label{sec:conclu}

In this article, we highlight the dynamic nature of political discourse on Twitter during the 2022 French presidential election. Our analysis begins by examining collective modes in the time series of political community activities, where the eigenvalue spectrum of filtered correlation matrices reveals periods of synchronized fluctuations. These large eigenvalues correspond to emergent collective behavior, especially noticeable before critical events like the election, indicating heightened coordination across political parties on key issues.

Building on this, we constructed correlation-based graphs from the activity changes, identifying central nodes (political parties and topics) that play pivotal roles in shaping the political landscape. These centralities reveal which actors and themes dominate the network of interactions, shedding light on the structural backbone of political discourse. However, while correlation captures mutual relationships, it does not resolve which node initiate the dynamical variation of activity that can cause a re-shape of the network. 

To address this limitation, our subsequent causal inference analysis reveals asymmetric patterns of influence among political communities. For example, parties such as Hidalgo and Jadot predominantly influence Mélanchon without being influenced in return, while Macron and Mélenchon exhibit a mutual, competitive shaping of discourse. Meanwhile, some parties like Zemmour act as influential initiators with broad impact across the spectrum, yet remain relatively uninfluenced by others, highlighting diverse strategic roles in political communication.

Sociologically, these findings reinforce the idea that political influence extends beyond electoral success to encompass narrative dominance and responsiveness to key external events, such as the COVID-19 pandemic and the Ukraine crisis. The varying degrees of receptiveness to influence suggest that some political actors remain insulated within their ideological spheres, while others are more permeable to cross-party dynamics. By mapping these interactions, our study offers empirical insight into how political actors engage with one another on social media, highlighting differences in responsiveness, influence, and topic centrality across the ideological spectrum. These findings suggest that influence within the political landscape extends beyond formal electoral outcomes and manifests through differential attention to key issues and asymmetries in discourse dynamics. This observation aligns with theoretical perspectives on political communication and media influence, such as \citep{bourdieu1994distinction} with the concept of the political field as structured by symbolic power relations, and complements deliberative democracy theory, which emphasizes the importance of discourse and inter-party engagement in shaping democratic outcomes \citep{habermas1984theory, dryzek2000deliberative}. 

Overall, the combination of collective mode detection, correlation-based centrality measures, and causal inference offers a powerful, data-driven framework to disentangle the complex, directional flows of political influence in digital environments. Future work could extend this approach by incorporating richer content analysis of the tweets themselves, examining how specific frames or rhetoric contribute to influence dynamics. Additionally, applying this methodology longitudinally across multiple election cycles could illuminate the evolution of political strategies and the resilience of influence networks. From a practical standpoint, these insights may assist campaign strategists and media analysts in monitoring shifts in political discourse and anticipating emerging coalitions or conflicts in real time.

\bibliographystyle{unsrtnat}

\bibliography{references}  






\section{Appendix}

\subsection{List of topics}
Here is the list of queries defining the topics use in our analysis. For each term of a query, a manual check has been made in the Politoscope database to guaranty that the tweets retrieved where actually related to the topic.

\textbf{Agriculture:} List(" agricult", " agro", " pac ", "alimentation", " engrais ", "circuits courts", "transition agricole", "surplus", " intrant", " bio ", "éleveurs", "monde rural", "fnsea", " ogm", "engrais", "produits agricoles", "pêche", "alimentation", "agroalimentaire", "glyphosate", "néocotimoides", "neocotimoides")

\textbf{Democracy:} List("penelope", "parlement", "trumpisation", " élus ", "corruption", " ena ", "non-cumul des mandats", "ve république", "6e république", "6eme république", "5e république", "vie politique", "open data", "données publiques", "peneloppe", "pénélope", "pénéloppe", "démocratie sociale", "cour des comptes", "proportionnelle", "parité", "assemblée nationale", "élections législatives", "fonction publique territoriale", "formation des fonctionnaires", "durée du quinquennat", " préfet ", "établissements publics", "révisions constitutionnelles", "réserve parlementaire", "statut pénal du chef de l'état", " scandale ", "inéligibilité", "fonction présidentielle", "e-démocratie", "penelope fillon", "assistant parlementaire", "conseil constitutionnel", "assemblée constituante", "république", "souveraineté", "référendum", "réduire le nombre de parlementaires", "souveraineté numérique")

\textbf{Economy:} List("entreprise", " pme", " pmi", "actionnaire", "relocalisation", "crédit impôt", "nationalisation", "commerçant", "plus-values", "délocalisation", "cotisations sociales", " scop", "niches fiscales", " gafa ", " cice ", "taux d'imposition", "retenue à la source", "crédit impôt-recherche", "impôt sur les sociétés", "services innovants", "marchés publics", "livret d'épargne industrie", "exportation", "impôt sur la fortune", "conseils d'administration", "small business act", " surplus", "capital-risque", "comités de rémunération", "représentants des salariés", "harmonisation fiscale", "complexité administrative", "reprise d'entreprises", "stock options", "start-ups", "pme européennes", "commandes publiques", "fiscalité européenne", "impôt minimal", "chiffre d'affaires mondial", "prélèvement à la source", "échanges commerciaux", "syndicalisation", "travailleurs handicapés", "commerce indépendant", "politique commerciale", "développement économique", "politique commerciale", "charges patronales", "tva anti-délocalisations", "allègements de charges", " cice ", "alléger les charges patronales", "centrales d'achat", "concurrence internationale", "exploitations familiales", "financiarisation", "travailleurs indépendants", "liberté syndicale", "austérité", "fisc", "impôt", "imposition", "cotisations sociales", "retenue à la source", "tranche supplémentaire", " isf", "prélèvement à la source", "cotisations chômage", "exonérations de cotisation sociale", "tva", "charges patronales", "tva", "csg", " taxe", "allègements de charges", "trésor public", "droits de mutation", "dette des états", "dette de l’états", "dette publique", "banque", "financi", "subprimes", "bancaire", "épargne", "spécul", "livret A", "agios", "taux d'usure", "emprunt", "agence de notation", "crédit à la consommation", "surendettement", "assurance-vie", "loi sapin", "banque centrale européenne", "euro-obligations", "pacte de responsabilité", "taxe mondiale sur les transactions financières", "taxe Tobim", "dépôts des particuliers", "crise bancaire", "crise financière", "evasion fiscale", "taxe d'habitation", "pouvoir d'achat", "impôt", "capital-risque", "paradis fiscaux", "fraude fiscale", "révolution fiscale", "fiscalité", "patriotisme économique", "loi sapin", "taxe foncière", "niches fiscales", "rentrées fiscales", "justice fiscale", "impôts", "optimisation fiscale")

\textbf{Employment:} List("travailleur", "emploi", "apprentissage", "chômage", "chômeur", " rsa", " smic", "flexisécurité", "formation professionnelle", "insertion professionnelle", "lycées professionnels", "formation initiale", "heures supplémentaires", "temps partiel", "contrat de professionnalisation", "conditions de travail", "emplois précaires", "enseignement professionnel et technologique", "compensation salariale", "39 heures", "charges sociales salariales", "employeurs", "filières professionnelles", "35 heures", "tutorat", "pôle emploi", "création d'entreprise", "salaire horaire", "négociation salariale", "exonérations de cotisation sociale", "insertion des jeunes", "parcours professionnels", "licenciement", "travail obligatoires", "jeune en alternance", "37 heures", "contrat de génération", "loi travail", "augmenter le salaire", "prime pour l'emploi", "code du travail", "formation des fonctionnaires", "dialogue social", "salaire des fonctionnaires", "loi el khomri", "sous-traitants", "devenir entrepreneurs", "souffrance au travail", "gel des traitements", "obligation pour eux d'accepter un emploi", "maitrise de la masse salariale", "non remplacement de un fonctionnaire sur deux", "suppressions de postes", "artisanat", "réduction du temps de travail", "RTT", "cice", "mobilité professionnelle", "âge légal de départ à la retraite", "retraite solidaire", "pénibilité", "salaire moyen", " cdi", "cdd", "délocalisations", "relocalisation", "revenu universel", "revenuuniversel", "contrat de travail", "retraite", "chômeurs", "assurance chômage", "code du travail", "plein emploi", "heures supplémentaires", "emplois précaires", "apprentissage", "pôle emploi", "loitravail", "augmenter le salaire")

\textbf{Ecology:} List(" écolo", "biodiversité", "protection de l’environnement", "pollutions", "diesel", "changement climatique", "fessenheim", "flamanville", "performance énergétique", "rénovation thermique", "isolation de l'habitat", "précarité énergétique", "gaz de schiste", "pesticides", "gaz à effet de serre", "ressources naturelles", "quotas de pêche", "agroécologique", "énergies marines renouvelables", "pêche illégale", "isolation de l'habitat", "forêts", "sylvicultures", "planification écologique", "transition écologique", " epr ", "sortir du nucléaire", "conversion écologique", "climat", "émissions de gaz à effet de serre", "agroécologique")

\textbf{Immigration:} List("migr", "immigr", "nationalité", "clandest", "droit de séjour", "schengen", "pays d'origine", "contrôle des frontières", "réadmission", "régularisations", "regroupement familial", "naturalisation", "reconduites à la frontière", "titres de séjour", "procédures d'expulsion", "séjour des étrangers", "libre circulation des personnes", "connaissance préalable de la langue française", "intégration des étrangers", "citoyenneté française", "droit du sol", "droit du sang", "islam", "musul", "muzz", "salaf", "burkini", "barbarie", "laïcité", "juif", "juive", "chrétien", "communautarisme", "multiculturalisme", "modèle républicain", "identité nationale", "civilisation française", "port du voile", "multiculturalisme", "grand remplacement", "génocide par substitution", "non-citizens", "illegal immigrants", "déchéance de nationalité", "réfugiés", "loi de 1905", "islamophobie", "frères musulmans", "salafisme", "bi-nationalité", "réadmission", "nationalité", "racines chrétiennes", "carte de séjour")

\textbf{Foreign politics:} List("europ", "souverain", " CETA", "TAFTA", "bruxelles", "russie", "russe", "pacte de stabilité", "directive européenne", "traité franco-allemand", "trump", "washington", "maison blanche", "moscou", "crise syrienne", "libye", " iran", "irak", "syrie", " OTAN", "putin", "poutine", "trumprussia", "turquie", "poutine", " ue ", "ukrain", "donbass", "marioupol", "zelensky", "kiev", "kremlin", "crimee", "crimée", "ukraine", "parlement européen", "bruxelles", "traités européens", "moyenne européenne", "commission européenne", "chine", "bachar al-assad", "frexit")

\textbf{Energetic politics:} List("électricité", "nucléaire", "énergies renouvelables", "éolien", "énergies marines renouvelables", "énergie renouvelable", "énergie solaire", " epr ", "flamanville", "charbon", "centrales thermiques", "diesel", "fessenheim", " éol", "flamanville", "performance énergétique", "fukushima", "rénovation thermique", "isolation de l'habitat", "précarité énergétique", "gaz de schiste", "énergies marines renouvelables", "isolation de l'habitat", "sources d'énergie", "transition énergétique", "conversion écologique", "énergies alternatives", "dépendance énergétique")

\textbf{Research and education:} List("lycée", " pisa ", "éducation", "apprentissage", "maternelle", "décrochage scolaire", "scolarité", "scolaire", "service civique", "savoirs fondamentaux", "méthodes pédagogiques", "enseignement professionnel et technologique", "établissements scolaires", "tutorat", "échec scolaire", "écoles d'ingénieurs", "financer leurs études", "banque de la jeunesse", "écoles de commerce", "e-learning", "éducation artistique", "enseignement primaire", "universit", "la recherche", "chercheur", "R\&D", "science", "scientifique", "enseignant", "professeur", "loi lru", "enseignants-chercheurs", "enseignement supérieur", "grandes écoles", "investissements d'avenir", "loi fioraso", "crédit impôt-recherche", "sciencespo")

\textbf{Healthcare: } List("santé", "médecin", "médic", "hôpit", "sécurité sociale", "autisme", "handicapés", "mdph", "désamiantage", "cancer", "assurance-maladie", "hébergement en établissement", "maintien à domicile", "soins d'urgence", "urgences sanitaires", "maladies", "perturbateurs endocriniens", "accès aux soins", "hospitalisation", "soins dispensés", "parcours de soins", "centre de soin", "malentendants", "malvoyants", "dépassements d'honoraires", "transports sanitaires", "arrêt de travail", "numerus clausus", "carte vitale", "perte d'autonomie", "ehpad", "agences de sécurité sanitaire et alimentaire", "tiers payant", "recours aux génériques", "lunettes", "handicap", "ehpad", "covid", "janssen", "vaccin","vaccins","vaccines","vaccine","vaccination", "passvac", "coronavirus", "hydroxychloroquine", "masques","masque", "covid-19","covid\_19","covid19","confinement","chloroquine", "Pfizer-BioNTech", "moderna","astrazeneca", "antivac", "antivax", "infirmières", "médecins", "pandémie", "carte vitale biométrique", "déserts médicaux", "quarantaine", "crise sanitaire", " orpea ", "hôpital public", "système de santé", "perturbateurs endocriniens", "obesité", "surpoids")

\textbf{Security :} List("terror", "attentat", "cyber-sécurité", "guerre civile", "surveillance", "délinqu", "espion", "loi taubira", "vidéo-surveillance", "vidéo-protection", "sécurité des français", "laxisme judiciaire", "prescription", "peines planchers", "remises de peine", "menace terroriste", "daech", "cyber-guerre", "cyber-attaques", "frontière", "menace terroriste", "délinquance des mineurs", "guerre cognitive", "guerre hybride", "ingérences étrangères", "ingérence étrangère")

\end{document}